\documentclass[sigconf]{acmart}

\usepackage{multirow}
\usepackage{lscape}
\usepackage{array}
\usepackage{enumitem}

\AtBeginDocument{%
  \providecommand\BibTeX{{%
    \normalfont B\kern-0.5em{\scshape i\kern-0.25em b}\kern-0.8em\TeX}}}

\copyrightyear{2022} 
\acmYear{2022} 
\setcopyright{acmcopyright}\acmConference[CHASE'22 ]{15th International
Conference on Cooperative and Human Aspects of Software Engineering}{May
21--29, 2022}{Pittsburgh, PA, USA}
\acmBooktitle{15th International Conference on Cooperative and Human Aspects of
Software Engineering (CHASE'22 ), May 21--29, 2022, Pittsburgh, PA, USA}
\acmPrice{15.00}
\acmDOI{10.1145/3528579.3529178}
\acmISBN{978-1-4503-9342-3/22/05}

\begin{document}

\title{Characterizing User Behaviors in Open-Source Software User Forums: An Empirical Study}

\author{Jazlyn Hellman}
\email{jazlyn.hellman@mail.mcgill.ca}
\affiliation{%
 \institution{McGill University}
 \city{Montreal}
 \state{Quebec}
 \country{Canada}}
 
 \author{Jiahao Chen}
\email{jiahao.chen@mail.mcgill.ca}
\affiliation{%
 \institution{McGill University}
 \city{Montreal}
 \state{Quebec}
 \country{Canada}}
 
 \author{Md. Sami Uddin}
\email{sami.uddin@mcgill.ca}
\affiliation{%
 \institution{McGill University}
 \city{Montreal}
 \state{Quebec}
 \country{Canada}}

\author{Jinghui Cheng}
\email{jinghui.cheng@polymtl.ca}
\affiliation{%
 \institution{Polytechnique Montreal}
 \city{Montreal}
 \state{Quebec}
 \country{Canada}}

\author{Jin L.C. Guo}
\email{jguo@cs.mcgill.ca}
\affiliation{%
 \institution{McGill University}
 \city{Montreal}
 \state{Quebec}
 \country{Canada}}

\begin{abstract}
User forums of Open Source Software (OSS) enable end-users to collaboratively discuss problems concerning the OSS applications. Despite decades of research on OSS, we know very little about how end-users engage with OSS communities on these forums, in particular, the challenges that hinder their continuous and meaningful participation in the OSS community. Many previous works are developer-centric and overlook the importance of end-user forums. As a result, end-users' expectations are seldom reflected in OSS development. To better understand user behaviors in OSS user forums, we carried out an empirical study analyzing about 1.3 million posts from user forums of four popular OSS applications: Zotero, Audacity, VLC, and RStudio. Through analyzing the contribution patterns of three common user types (end-users, developers, and organizers), we observed that end-users not only initiated most of the threads (above 96\% of threads in three projects, 86\% in the other), but also acted as the significant contributors for responding to other users' posts, even though they tended to lack confidence in their activities as indicated by psycho-linguistic analyses. Moreover, we found end-users more open, reflecting a more positive emotion in communication than organizers and developers in the forums. Our work contributes new knowledge about end-users' activities and behaviors in OSS user forums that the vital OSS stakeholders can leverage to improve end-user engagement in the OSS development process.

\end{abstract}

\ccsdesc[500]{Human-centered computing~Empirical studies in collaborative and social computing}
\ccsdesc[500]{Software and its engineering~Open source model}
\ccsdesc[500]{General and reference~Empirical studies}


\keywords{Open-source software, user forums, linguistic analysis}

\maketitle





\section{Introduction}
The impact of Open Source Software (OSS) is increasing rapidly. As OSS grow, they amass an increasingly large and diverse user base in addition to a variety of applications across different domains, such as scientific computing, programming, media editing, and content management. Successful development and maintenance of OSS involves frequent and direct communication among different stakeholders, such as developers, maintainers, project owners, contributors, designers, and end-users, to facilitate an understanding of each stakeholder's unique needs and challenges. These efforts ultimately serve as the main drive to build a thriving OSS community with diverse groups engaging in the development and maintenance process.

One such communication channel for OSS is the user forums where the stakeholders of OSS may come to engage with the larger community. In particular, end-users of an OSS rely on these user forums to discuss usage issues, collaboratively find solutions to their problems, request new features, and more. Despite forums' longstanding use, we know very little about the practices and activities of different stakeholders in these user forums, particularly what challenges prevent end-users from engaging frequently and meaningfully with the OSS community. 

Early research on OSS development is primarily concerned with the stakeholders who participate in the code contribution of the projects. The investigations related to forums were limited to developer forums and tasks mostly concerning developers and organizers of OSS, such as new developers' on-boarding~
\cite{fagerholm2014role}, contributors social network~\cite{8812044}, discussion instrumentation~\cite{argulens}, and engaging organizers to OSS~\cite{LinkGJP}. A handful of studies on end-users only focused on certain behavior of the power users, such as bug reporting~\cite{10.1145/1753326.1753576}. The communication with the mass end-users, however, has largely been ignored in previous OSS studies.
By analyzing the end-users' activities and communication behaviors with developers, organizers and other users in forums, we can find insights into their actions, opinions, and mental and emotional states when using and discussing the OSS. These insights are prerequisites to support the efforts of OSS communities to improve the usability~\cite{wang_how_2020} and accessibility of the software~\cite{heron2013open}, as well as the sustainability of the OSS communities themselves~\cite{benbya2016successful}.  


In this work, we aim to characterize the current landscape of OSS end-user forums to identify the challenges and opportunities for end-user engagement and community building. Towards this end, we carried out an empirical study on user forums of four popular OSS applications: Zotero, Audacity, VLC, and RStudio. We analyzed a total of about 1.3 million posts created by more than 200,000 participating members in those four user forums. Our analysis sought to answer the following research questions:

\begin{enumerate}[noitemsep, leftmargin=*, itemindent=0pt, label={\textbf{RQ{{\arabic*}}}:}]
    \item What are the characteristics (e.g., length, duration, response time) of the discussion threads in OSS user forums?
    \item How do different roles participate in the OSS user forums, including end-users, developers, and organizers?
    \begin{enumerate}[noitemsep, leftmargin=24pt, itemindent=0pt, label={\textbf{RQ2.{{\arabic*}}}:}]
        \item  How frequently does each role perform different activities related to the discussion threads?
        \item How do the content of their posts differ in terms of linguistic features that capture their cognitive processes, emotions, motivation, and power dynamics?
    \end{enumerate}
\end{enumerate}

To answer these research questions, we analyzed the forum posts separated by the categories of forum users' roles and the types of activities these forum users performed. In particular, we focused on the frequency as well as the breadth of interactions among users within the forums (e.g., response time and user tenure). Moreover, we conducted psycho-linguistic analyses on the forum contents to understand the emotional and mental states of interactions among users (e.g., level of confidence, cognitive states) in OSS forums. 

Results revealed that organizers, developers, and end-users contributed to three types of activities (i.e., first post to initiate a thread, reply from other users, and reply from an original poster) in OSS user forums with distinct patterns. As expected, end-users substantially contributed to the OSS forums by initiating discussions. End-users also actively participated in conversations, in addition to organizers and developers, by responding to other users' posts. Although end-users generally lacked confidence in their communication compared to the other two roles, our results indicate the presence of a large pool of active and open-minded end-users who can contribute to improving the usability of OSS by providing relevant feedback. 

Overall, this study provides three main contributions. First, we contribute a novel set of characteristics which reveal that besides discussing problems concerning OSS applications, end-users of OSS make significant contributions to the OSS community by providing feedback to other users. Second, we show that despite OSS end-users having fairly neutral tones when conversing, they bring a more positive attitude and openness to the user forums when compared to the organizers and developers, even though the end-users exhibit a lack of confidence while interacting. Third, we present an empirical evidence-based characterization of user behaviors in OSS user forums that can be leveraged to increase end-user engagement and collaboration in the OSS community and eventually contribute to OSS improvement.

\section{Background \& Related Work}

    \subsection{OSS Usability and Forums}
Past research has established that in OSS, end-users discuss their issues in an ad-hoc manner on forums, when they are present, or in a repository's issue tracking system such as on GitHub~\cite{FoSS_usability, argulens, cheng_guo_uxissues}. Cheng and Guo~\cite{cheng_guo_uxissues} found that usability specific issue threads are lengthy and contain over-generalized assumptions, where usability refers to attributes which determine ease, error-prevention, efficiency, and pleasantness for an end-user when interacting with a software~\cite{cheng_guo_uxissues, nielsen_book}. Wang et al.~\cite{wang_how_2020} advocated for the need for a user-centric and inclusive mindset amongst OSS practitioners. Hellman et al.~\cite{hellman_facilitating_2021} identified inclusiveness and learnability of issue tracking as barriers for non-technical OSS end-users to collaborate on a project in addition to OSS designers and developers struggling to manage multiple sources for end-user engagement.  Most prior work on OSS forums and discussions has focused on developer forums, discussions, or mailing lists but not the end-user specific forums~\cite{4228660, 10.1145/3196369.3196372, 10.1145/3475716.3475775, oss-emails-speech-act-2015}. 

 \subsection{Forums and Text Analysis}
Prior efforts have been made to establish taxonomies for online communications so that automatic classifiers may aid efforts in these communications~\cite{oss-emails-speech-act-2015,twitter_dialogue_2017, twitter_taxonomy_dialgoue_2019, kim_wang_baldwin, jeong_semi-supervised_2009}. Ivanovic \cite{ivanovic_2005} established a taxonomy to model and detect dialogue acts (i.e., statement, thanking, open-answer, etc.) in instant messaging. 
Specifically in regards to online forums,  Zhang et al.~\cite{health_forum_user_intents_2014} propose a taxonomy for user intents in online health forums and 
Zhang et al. \cite{zhang_characterizing_nodate} characterized online discussion comments into discourse acts (e.g., question, answer, announcement, etc.). 
Software requirements extraction efforts have attempted to leverage online forums and discussions from users. Stade et al. \cite{stade-requirment-tool-2017} built an end-user forum for requirements extraction, but found a forum alone is insufficient, while Kanchev et al. \cite{canary-2017} used high-level query language to automatically extract requirements from online discussions. In regards to technical OSS forums, prior efforts have explored methods to understand forums and its users. Nugroho et al.~\cite{nugroho_how_2021} investigated project-specific forum threads in the Eclipse ecosystem to understand participation, content, and sentiment. Wang et al. \cite{wang-incentive-2018} explored ways to improve incentive systems for fast answers in technical Q\&A websites. 
    
\subsection{Linguistic Inquiry and Word Count}
One established text analysis method is the Linguistic Inquiry and Word Count (LIWC) tool which uses word counting methods for analyzing the psychological meaning through specific categories (e.g., articles, positive emotions, power, etc.) of words present in dialogue and texts \cite{LIWC2015-paper, Tausczik2009}. LIWC utilizes an extensive dictionary, which has seen multiple developments, with the latest version, LIWC2015, consisting of about 90 categories with almost 6,400 words, word stems, and select emoticons in addition to its processing component \cite{LIWC2015-paper, Tausczik2009}. The LIWC application takes a text input (of any length) and for each word in the text, a comparison is made to the dictionary. When a word is present in the dictionary, the category containing the word is incremented (some words may belong to multiple categories) and a value is calculated for each category to represent the percentage of the total original text \cite{Tausczik2009}. 
    
In practice, LIWC has been applied to various fields to understand relationships between psychological meaning and language used to communicate. While extensive research has validated LIWC's accuracy in a variety of domains, LIWC is constantly improving and it remains that accuracy will vary by specific tasks and datasets~\cite{Tausczik2009,LIWC2015-paper,bantum2009evaluating}.
Some previous studies that used LIWC include: predicting relationship conflict interactions~\cite{10.1145/3461615.3485423}, understanding help-seeking behavior on social media~\cite{10.1145/2531602.2531720}, measuring linguistic style accommodation on social media \cite{mark-my-words}, and examining emotions of different forum participants in mental distress on discussion forums \cite{10.1145/1940761.1940914}.
Specifically in the context of OSS, LIWC has successfully been utilized to examine types of developer personalities to understand markers of release success, the likelihood to become project contributors, and the implications of inferring personality from psycho-linguistic methods~\cite{4228660, 10.1145/3196369.3196372, 10.1145/3475716.3475775}. 
For our purpose of understanding the communication patterns in OSS user forums, we are applying LIWC to extract meaningful psycho-linguistic characteristics in OSS forum posts. Section~\ref{sec:liwc-method} details our category selection and methodology for inspection.

\section{Methods}

\subsection{OSS Projects Selection}
To answer our research questions, we scope the projects under investigation with three criteria. \textit{First, each project needs have an active user forum.} The primary goal of our study was to investigate how end-users participate in discussions and what facilitates or challenges them in the existing platform. User forums are the primary channel to enable end-users to communicate with other community members. \textit{Second, the OSS should have a Graphical User Interface (GUI).} We chose OSS with a GUI because their user base is normally more diverse in terms of programming backgrounds. The gap between the end-users and developers can potentially impact how they communicate problems related to the software. An analysis on these projects is especially informative and important towards OSS usability and inclusiveness. \textit{Last, the project manages their code base on GitHub.} We would be able to identify the contributors of each project and cross-reference if they possess any role in the user forum. Following those criteria, we selected four popular OSS projects for analyses: Zotero, Audacity, VLC, and RStudio:

\noindent $\bullet$ \textbf{Zotero} (www.zotero.org) is a popular reference management software that helps users collect, maintain, and cite reference materials. 
Zotero provides a basic forum\footnote{https://forums.zotero.org/discussions} for the users to post their questions or comment on other forum users' posts. All threads other than the announcements from the administrators are ordered chronologically with the latest on top. Forum users can also use the search button to initialize a query based search on the existing posts.

\noindent $\bullet$ \textbf{Audacity} (www.audacityteam.org) is a multi-track audio recorder and editing application. 
The user forum for Audacity\footnote{https://forum.audacityteam.org} is divided into several modules, including `audacity help forum', `feedback and discussion forum', `special interest group', and `programming and development'; each module is further split into sub-forums for concrete topics.
The users can search the forum with textual queries or browse the unanswered topics through quick links.

\noindent $\bullet$ \textbf{VLC} (www.videolan.org) is a cross-platform multimedia player that supports various media formats and can be used as a streaming server. The design of VLC user forum\footnote{https://forum.videolan.org} greatly resembles Audacity with only a slight difference in some of the concrete sub-forums. 

\noindent $\bullet$ \textbf{RStudio} (www.rstudio.com) is an Integrated Development Environment, specially designed to facilitate R programming. It is frequently used for data science and scientific research. The design of the RStudio user forum\footnote{https://community.rstudio.com} is a combination of the design of the other forums discussed above. All threads are listed chronologically with pinned threads on top. Each thread can be assigned to one category and one or more tags to enable the forum users to search and browse them with such meta-data. This user forum also integrates additional functions to support engagement, such as the like buttons for each posts and new topics recommendations.

\subsection{Data Collection}
\label{subsec:data_collection}
We first extracted the threads of posts from each forum using Scrapy\footnote{www.scrapy.org}. The range of posts collected is from the inception of respective forums till October, 2021. Concretely, we extract the textual content of each post, the topic it belongs to, poster authors' user names and user type, and the timestamp of the post. Table~\ref{tab:data_all} presents an overview of the forum data we collected during this step. Overall, we compiled a total of 282,946 threads from the four OSS forums. These thread included over 1.29 million posts made by 207,654 members. These four forums represent various forum sizes and activity levels. Among them, VLC had the highest number of members ($N=94,210$), while RStudio had the lowest ($N=21,790$).

Using the GitHub API, we then collected the GitHub user name for all the contributors for each OSS. Initial manual inspection indicates that developers tended to use the same username in the user forums as on GitHub. Therefore, by comparing the user names on these two sites, we labeled each forum user if their role is likely to be a developer. 

\begin{table}[t]
  \caption{Total Number of members, threads and posts extracted for analysis.}
  \label{tab:data_all}
  \begin{tabular}{cccc}
    \toprule
    OSS&Members&Threads&Posts\\
    \midrule
    \ Zotero & 51,377 & 70,854 & 395,220\\		
    
    \ VLC & 94,210 & 120,364 & 418,895\\
    \ Audacity & 40,277 & 48,449 & 313,526\\
    \ RStudio & 21,790 & 42,379 & 167,370 \\
    \midrule
    \ Total & 207,654 & 282,946 & 1,295,011 \\

  \bottomrule
\end{tabular}
\end{table}


\subsection{Data Analysis}


\subsubsection{Analysis of forum discussion threads}
In OSS forums, threads enable topic specific discussions. A thread begins with a user creating a post, and then members of that OSS community joining the thread by responding to its posts. We analyzed the progression of threads in four OSS forums in terms of thread length and time span. Thread length acts as a proxy for the complexity of the topic under discussion and time span can reflect how active and responsive the community is. Early study on developer discussion forums have adopted similar metrics to measure community evolution~\cite{MoutidisSO}. 

To understand user behaviors in OSS forums, we also analyzed the cases when the threads lacked response from the community. In particular, we define \textbf{Ignored Threads} as the ones that did not receive comments from members other than the original post author, and define \textbf{Abandoned Threads} as those that never received a follow-up comment from the original author. We calculated the percentages of those threads in each forum to understand the responsiveness of thread authors and the OSS communities.


\subsubsection{Analysis of forum participants' roles and their activities}
\label{subsubsec:role_method}

We classified all forum users into three categories: Organizers, Developers, and End-Users considering their roles in the OSS community.
\textbf{Organizers (ORG)} manage and maintain the forum's integrity and quality. They are assigned administrative privilege on forums and perform duties such as approving, curating, or even suspending posts made by other forum users. They can also directly contribute to the discussion. This role is named differently in each forum and can be further divided to sub-roles. For example, it is called Admin in Zotero and Site Administrator in VLC. For Audacity, such role includes Site Admin, Forum Staff, Quality Assurance, etc. In RStudio forum, it includes Sustainer and RStudio Employee. To avoid confusion we use the term Organizers to refer to them. 
\textbf{Developers (DEV)} contribute to developing the respective OSS. Specifically, we considered forum users who have code contributions to the repositories of the respective OSS in GitHub as DEV. As described in Section~\ref{subsec:data_collection}, this is achieved through matching the GitHub usernames of contributors of an OSS to the usernames of the associated forum users for the OSS. We excluded the forum users who have already been categorized to an ORG role during this step to avoid multiple labels.
We consider the forum users who were not categorized as DEV or ORG as \textbf{End-Users (EU)} of the OSS.

Along the axis of OSS community roles, we first considered concrete activity types and their frequencies related to posting (1) Initialize a thread, i.e., create the first post of a thread in an OSS forum; (2) Follow up the thread initialized by themselves by writing reply post; (3) Contribute to threads that other users inside a forum initialized. This analysis aimed to understand the overall communication pattern of the OSS forum users. We then compared the user tenure, in terms of active days in the forum, of the three categories of forum users.

\subsubsection{Analysis of linguistic features of different roles}
\label{sec:liwc-method}

\begin{table*}[t]
    \centering
    \small
    \caption{Summary of LIWC categories for analysis. The first four categories are unique summary calculations and the remaining categories are dictionary-based measurement for psychological processes~\cite{Tausczik2009, LIWC2015-paper}.}

   \resizebox{\textwidth}{!}{%
    \begin{tabular}{p{.07\linewidth} p{.42\linewidth} p{.44\linewidth}}
        \toprule
         \textbf{Category} & \textbf{Explanation} & \textbf{Justification} \\
         \hline
         Analytic & Degree of formal, logical, and hierarchical thinking exhibited. A higher score refers to the presence of more analytical contents \cite{10.1371/journal.pone.0115844}.  & OSS forums experience complex discussions between users, we select this category to investigate how different user groups use logical and analytical expressions.  \\ \hline
         Authentic & How much the text displays an honest, personal, or disclosing voice with a lower number indicating a more guarded or distanced form of speech~\cite{doi:10.1177/0146167203029005010}. & Users often disclose issues or experiences to get feedback; we select this category to analyze how users go about this process and how this reflects in their communication styles.  \\ \hline
         Clout & The amount of expertise and confidence a forum user possesses (reflected in their posts), with a lower score indicating a more tentative, humble, or anxious voice~\cite{doi:10.1177/0261927X13502654}. & To understand the level of confidence forum users exhibit while participating in the forum.\\\hline
         Tone &  The emotional tone of a text; a high score represents more positive and upbeat styles, while a lower score revealing anxiety, sadness, or hostility 
         \cite{doi:10.1111/j.0956-7976.2004.00741.x}. & OSS forums are a place for people to go for help when they experience problems, inspecting `Tone' will allow us to understand how different user groups.   \\\hline
         Cognitive Processes & Depth and complexity in which forum users are processing and interpreting information \cite{Tausczik2009}. & 
         This category will allow us to analyze how forum users think and process information.  \\\hline
         Affective Processes & Indicates how users express emotion in their posts. It also reveals how an individual is coping with an event \cite{Tausczik2009}. & Identify how users express emotions to events that either lead them to use the forum or happened while using the forum.
        \\\hline
         Drives & Degree of the needs, motives, drives, risks, rewards, and power references mentioned by the forum users \cite{LIWC2015-operator}. &  
         We analyze this category to investigate the amount of drives and motivations expressed by forum users. \\\bottomrule
    \end{tabular}%
    }
    \label{tab:LIWC-categories}
\end{table*}

To investigate how forum users' roles are correlated with their intentions, emotions, confidence, and other psychological states, we further conducted an linguistic analysis on the content of their posts. 
In particular, we selected seven categories from the LIWC2015 dictionary \cite{LIWC2015-paper}. These categories were identified as most relevant to the context of OSS user forums and communication behaviors. Table~\ref{tab:LIWC-categories} summarizes these categories and our rationale of choosing them.

To calculate the LIWC scores for these categories, we first isolated the individual natural language posts from the scrapped data and tagged each post with the user role. We then directly used LIWC to process the text documents; LIWC performs basic NLP preprocessing steps such as tokenization, handling capitalization, and stemming~\cite{LIWC2015-operator,LIWC2015-paper}. 
To understand how each role scores differently on each LIWC category for each forum, we considered the seven LIWC categories as the dependent variables and the categories of forum users as the independent variable (i.e. end-user, developer, and organizer) and conducted a statistical analysis.
We first performed Shapiro-Wilk tests on our dataset and determined that our sample does not follow a normal distribution on all dependent variables. Levene's tests also revealed that all dependent variables had unequal variances on the independent groups. 
Then we performed the Kruskal Wallis test on our dataset, followed by a pair-wise post-hoc test (Bonferroni corrected Mann-Whitney U) for all significant results. We finally calculated the effect size for each significant finding.

\section{Characterizing Forum Threads (RQ1)}
\subsection{Thread Life Span Analysis}

\begin{figure}[t]
\centering
\includegraphics[width=0.8\linewidth]{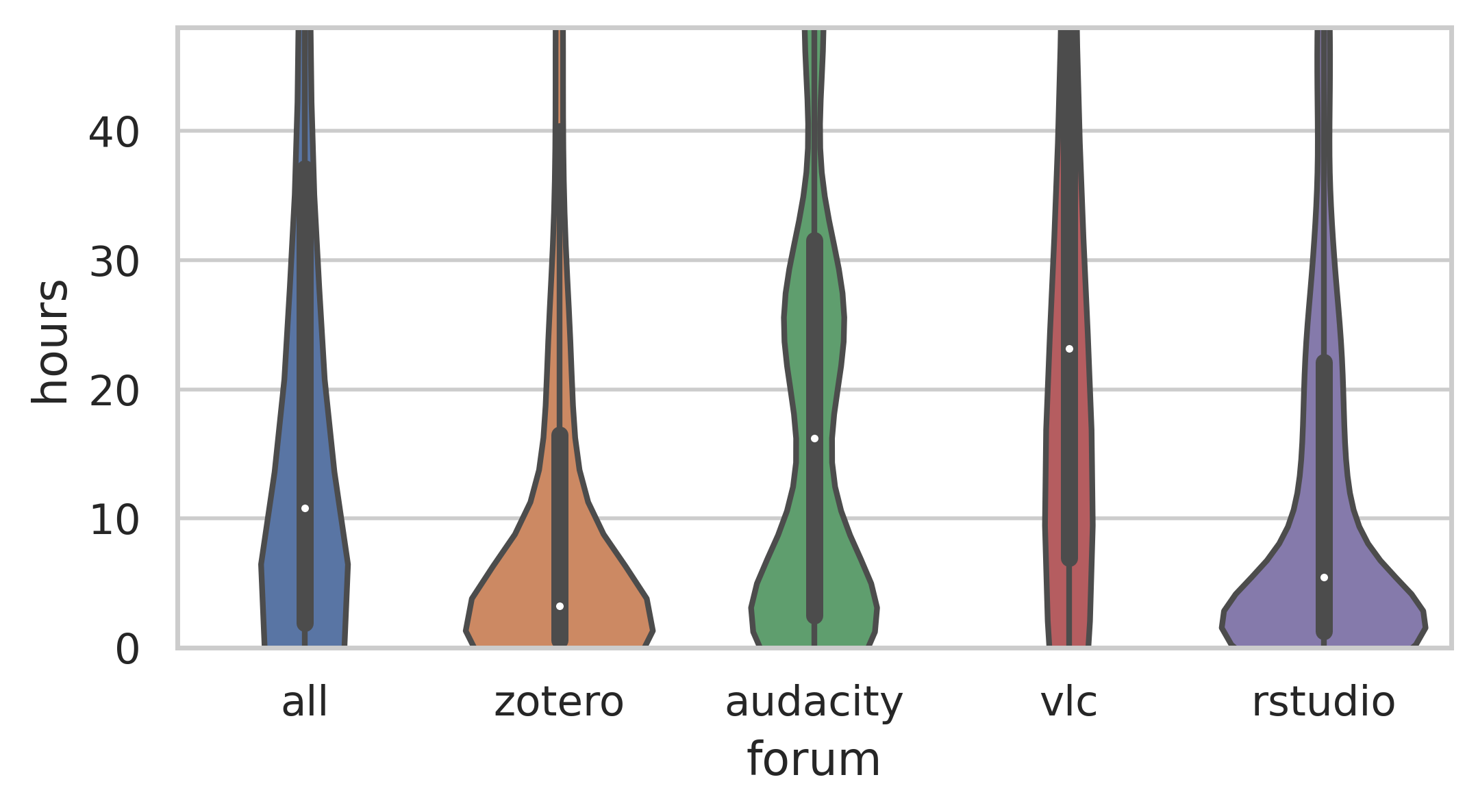}
\caption{Distribution of time until the first Non-Original Poster's Response in a thread (with outliers removed based on the $1.5*IQR$ criterion).}
\label{fig:first_NOPR}
\Description{text.}
\end{figure}

For RQ1, we began with examining how responsive the community was to new threads in the forum by analyzing how long it took for a thread to receive the first response from the forums users who were not the author of the initial post. There were in total 57,230 threads in our dataset (7,661 in Zotero, 2,007 in Audacity, 35,913 in VLC, and 11,649 in RStudio) that did not receive any response from other forum users than the original poster or did not receive any response at all; we leave the analysis on those threads in Section~\ref{subsec:ignored_abandoned}. We next removed the outliers based on the 1.5 IQR criterion; a total of 38,954 threads were removed in this process, including 12,644 from the Zotero forum, 7,181 from the Audacity forum, 14,291 from the VLC forum, and 4,838 from the RStudio forum. Figure~\ref{fig:first_NOPR} shows the distribution of the feedback time of threads in user forums with outliers removed. Across all forums, the median community response time was 10.83 hours ($IQR=35.15$). Zotero has the shortest median community response time of 3.22 hours ($IQR=15.92$) while VLC has the longest median community response time of 23.15 hours ($IQR=73.53$). Interestingly, the community response time in the Audacity forum follows a multi-peak distribution of roughly 24-hour intervals, indicating that the responses tended to happen during certain times of the day.

\begin{figure}[t]
\centering
\includegraphics[width=0.8\linewidth]{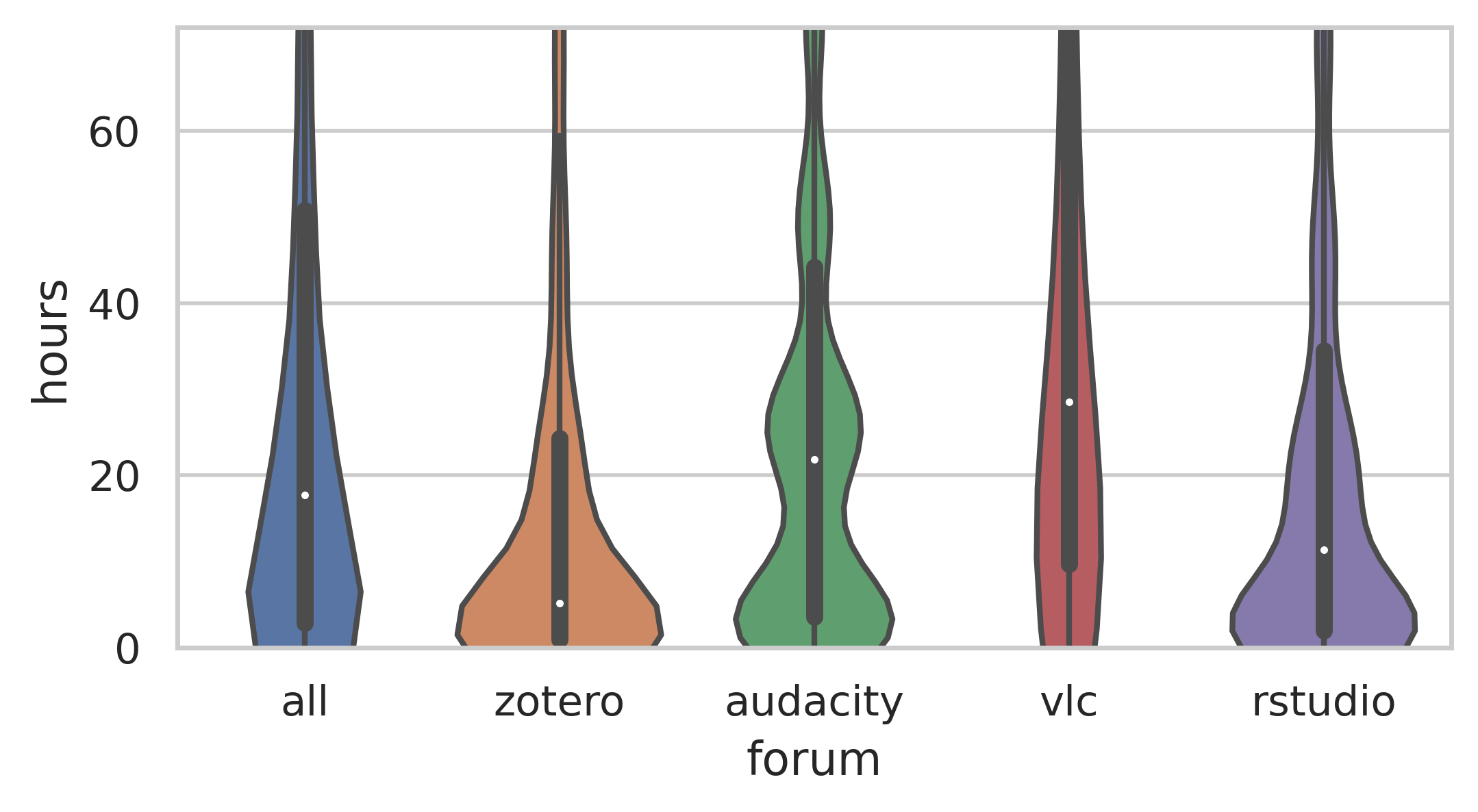}
\vspace{-0.2cm}
\caption{Distribution of threads life span (i.e. from the first post to the latest response) in the four OSS forums (with outliers removed based on the $1.5*IQR$ criterion).}
\label{fig:time_span}
\vspace{-0.2cm}
\end{figure}

We then calculated the life span of a discussion thread as the number of days elapsed from the initial post to the latest response in a thread. During this analysis, we ignored in total 47,824 threads (Zotero: 6,366, Audacity: 1,573, VLC: 29,985,  and RStudio: 9,900) which did not receive any response. We also removed the outliers on the remaining data based on the 1.5 IQR criterion; a total of 39,769 threads were removed in this process, including 12,749 from the Zotero forum, 7,346 from the Audacity forum, 14,928 from the VLC forum, and 4,746 from the RStudio forum. Figure~\ref{fig:time_span} shows the distribution of threads' life span in the four forums with outliers removed. The median thread life span for all four OSS forums was 17.72 hours ($IQR=47.88$). Among the four project, Zotero has the shortest median thread life span of 5.13 hours ($IQR=23.42$), followed by RStudio ($median=11.40$~hours, $IQR=32.48$). Audacity ($median=21.83$~hours, $IQR=40.62$) and VLC ($median=28.50$~hours, $IQR=90.74$) tended to have longer thread life spans.

\subsection{Thread Length Analysis}
To explore the density of discussions in the OSS forums, we calculated the number of posts found in each thread. Our results indicated a consistent pattern in threads' lengths across the four OSS forums, with a median length of 3 posts ($IQR=3.0$). Similarly, we found that the number of unique forum users who participated in the discussion threads was also consistent across the four forums, with a median of 2 users ($IQR=1.0$). Combining these results with the life span analysis, we conclude that most of the discussions in OSS end-user oriented forums are short-lived with a small amount of information exchange among a few community members. At the same time, however, thread lengths and unique user numbers in each thread had wide distributions, with a maximum of 1,928 posts that involved 478 users (from the Zotero forum). One reason behind the length of threads could be the complexity of topics discussed in the threads. For example, in the Zotero forum, a lengthy discussion (32 posts) took place on creating a reference entry from a PDF file, and due to huge demand, that feature was included in Zotero.

\subsection{Ignored and Abandoned Threads}
\label{subsec:ignored_abandoned}
We saw that a significant number of threads in OSS forums failed to reach a decisive resolved state. Those threads mostly fell into two categories; i.e., they either did not receive any response from the community (\textbf{Ignored Threads}) or the author of the initial posts did not follow up to the threads afterwards (\textbf{Abandoned Threads}). 

We found that the percentages of ignored threads differed greatly across the four OSS forums. The lowest percentage was found in the Audacity forum, with 4.14\%, followed by Zotero (10.81\%). The VLC and RStudio forums tended to have a higher percentage of ignored threads (31.52\% and 27.53\%, respectively). These results may have reflected how various types of community members respond to the posted topics differently across the four forums. We leave this analysis in Section~\ref{sec:participation_of_roles}.

Surprisingly, almost half (48.77\%) of the threads posted in the four forums were abandoned. The percentages of abandoned threads in the individual forums did not differ much, with the highest rate found in the VLC forum (56.14\%), followed by RStudio (47.25\%), Audacity (43.64\%), and Zotero (41.32\%). We hypothesize that this is due to the overall culture and the general behavior of forums' users. Even when the topic from the initial posts have already been fully discussed and addressed by the community, the original authors were not aware of the necessity or sufficiently motivated to return to the thread to confirm their status or express their gratitude.

\section{Participation of Various Roles (RQ2)}
\label{sec:participation_of_roles}
Following the method described in Section~\ref{subsubsec:role_method}, we divided the roles of forum users into three categories, i.e. Organizers (ORG), Developers (Dev), and End-Users (EU). Table \ref{tab:dist_users} presents a summary of the distributions of those roles in the four OSS forums. We also calculated the End-Users-to-Non-End-Users ratio, representing the hypothetical workload if a topic initialized by the end-users requires the organizers or developers to resolve. Audacity represents the highest ratio with 3008 EU to one Non-EU. The short respond time and low ignored thread rate for Audacity seems to suggest the community is coping with the situation well. In this section, we take a close look at the actions taken by different roles in the forums and the content of their posts to understand how they participate and collaborate in their communities.

\begin{table}[t]
  \caption{The distribution of different forum user categories for each OSS project.}
  \small
  \label{tab:dist_users}
  \begin{tabular}{ccccc}
    \toprule
    Project Name&\#ORG & \#DEV  &\#EU & EU/Non-EU Ratio\\
    \midrule
    \ Zotero & 2 &  24 & 51,351 & 1911\\		
    \ VLC & 9 & 37 & 94,164 & 2035\\
    \ Audacity & 8 & 5 & 40,264 & 3008\\
    \ RStudio & 28 & 26 & 21,736 & 396\\
  \bottomrule
\end{tabular}
\end{table}

\subsection{Analysis of Participation Type (RQ2.1)}
\label{sec:frequency-participation-results}

\subsubsection{Post Type Distribution}
In our analysis, we considered three types of participation of the forum users: (1) posting to start a thread (\textit{Initial Post}), (2) following up to a thread that is initiated by themselves (\textit{Replies by Original Poster}), and (3) respond to a thread initiated by someone else (\textit{Replies by Others}).
Among all the posts across the four forums, 22.74\% were \textit{initial posts}, 25.31\% were \textit{replies by original poster}, and 51.94\% were \textit{replies by others}. Both VLC and RStudio forums contained higher percentages of \textit{initial posts} (27.46\% and 25.08\% respectively). In contrast, Zotero and Audacity forums had lower percentages of \textit{initial posts} (18.21\% and 20.21\% respectively) but higher ratio of \textit{replies by others} (56.33\% and 54.78\% respectively), indicating a more engaged information exchange among their members.
Lastly, the fact that about half of all posts for each forum were \textit{replies by others} suggests that, regardless of the OSS project, members of the OSS forums were actively engaged in discussions.

\begin{figure}[t]
 \centering
 \includegraphics[width=\linewidth]{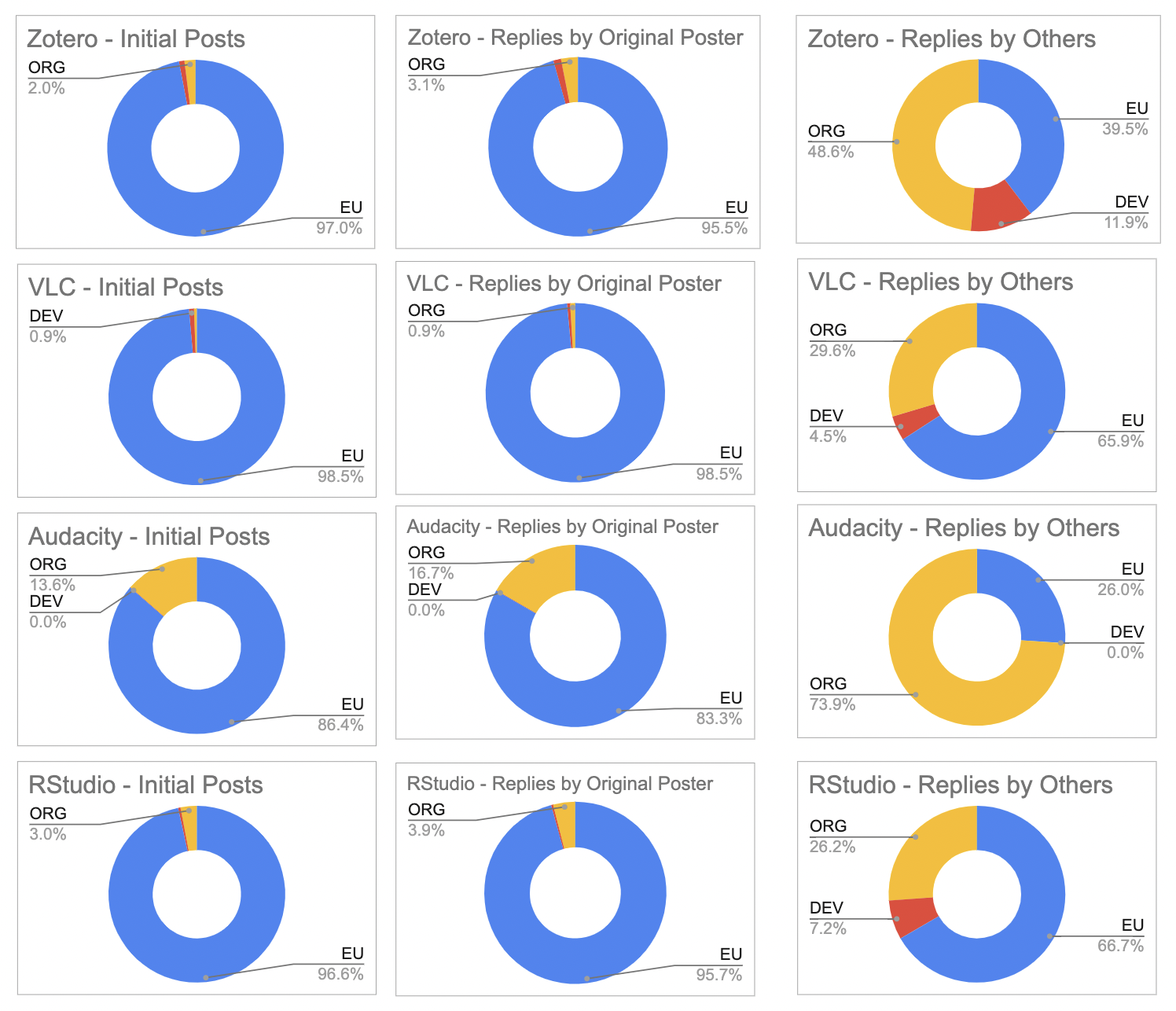}
 \caption{Analysis of users and posts types in four forums.}
 \label{fig:user_vs_post}
 \Description{text.}
\end{figure}

\subsubsection{Participation of different forum user roles}
We calculated the frequencies of each category of forum users creating posts for each post type, summarized in  Figure~\ref{fig:user_vs_post}.
As depicted in the left column of Figure~\ref{fig:user_vs_post}, end-users made the majority of the \textit{initial posts} across all four OSS forums. Consequently, they were also the ones who contributed to the majority of the \textit{replies by the original poster} (middle column of Figure~\ref{fig:user_vs_post}). Organizers and developers only occasionally initialized a topic, normally related to announcements of the software product, development progress, and forum rules.

With respect to replying to others' posts, the distributions of role categories are more diverse among the four forums (see Figure~\ref{fig:user_vs_post}, right column). In the Audacity forum, most of the responses were provided by a handful of forum organizers. For Zotero, forum organizers and developers without forum privileges together contributed to 62.4\% of the replies to others' post. The active responses from the organizers and developers may have contributed to the lower rate of ignored issues in the Audacity and Zotero forums that we found before (see Section~\ref{subsec:ignored_abandoned}). Meanwhile, a large number of end-users participated in discussions initiated by others, especially in the VLC and RStudio forums (64.5\% and 75.8\%, respectively). These results indicate that overall end-users play a critical role in OSS communities in providing feedback and support to other forum users, besides seeking resolution to their own issues.

\subsubsection{Tenure of forum users}

\begin{figure}[ht]
 \centering
 \includegraphics[width=0.9\linewidth]{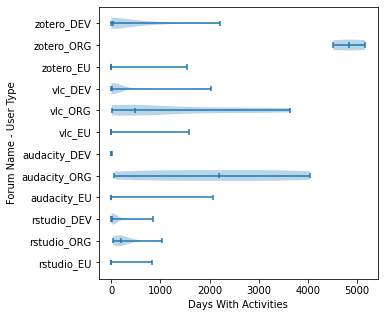}
 \caption{Analysis of active days in forums by roles.}
 \label{fig:active_days}
 \Description{text.}
\end{figure}

We used the tenure of forum users (i.e., time between the first and the last posts of the forum users) to indicate how long the forum users have participated in the OSS community. Figure~\ref{fig:active_days} summarizes this information for end-users, developers, and organizers across the four OSS forums. As expected, end-users generally had shorter tenures than developers and organizers across the four forums. However, there was a big variance in the duration of end-users' activities. In all four forums, there were some end-users who started their contribution since the beginning of the forum and were still active at the time of our data collection. 

\subsection{Analysis of Post Contents (RQ2.2)}
\label{sec:liwc-results}

To gain further insight into how the three OSS roles approached communications in the user forums, we performed an analysis on the textual content of the posts using linguistic features. In this analysis, we considered three independent groups, i.e., end-users (EU), developers (DEV), and organizers (ORG) and seven dependent variables, i.e., analytic, clout, tone, authentic, cognitive process, affective process, and drives calculated using LIWC. Kruskal Wallis Tests revealed statistically significant results among the three independent groups on all LIWC categories across all forums ($p<0.001$), excluding the cognitive process category for VLC ($p>0.05$). The results for the post-hoc Mann-Whitney U Test and the effect sizes can be seen in Table~\ref{tab:post-hoc-results}.

\begin{table*}[t]
\caption{Summary of relationships between psycho-linguistic behaviors  and participant groups. (*** p<0.001, ** p<0.01, * p<0.05; $\epsilon2$: negligible<0.01, weak<0.04, moderate<0.16, relatively strong<0.36, strong<0.64 \cite{rea2014designing}.)}

\centering
\resizebox{\textwidth}{!}{%
\begin{tabular}{ccccccccccccccc}
\multicolumn{3}{c}{} & \multicolumn{3}{c}{Zotero} & \multicolumn{3}{c}{VLC} & \multicolumn{3}{c}{Audacity} & \multicolumn{3}{c}{RStudio} \\ \hline
LIWC Category & Group (I) & \multicolumn{1}{c|}{Group (J)} & Median (I) & Diff(I-J) & \multicolumn{1}{c|}{$\epsilon2$} & Median (I) & Diff(I-J) & \multicolumn{1}{c|}{$\epsilon2$} & Median (I) & Diff(I-J) & \multicolumn{1}{c|}{$\epsilon2$} & Median (I) & Diff(I-J) & $\epsilon2$ \\ \bottomrule
\multirow{6}{*}{Analytic} & \multirow{2}{*}{End-user} & \multicolumn{1}{c|}{Developer} & \multirow{2}{*}{70.69} & -7.61 *** & \multicolumn{1}{c|}{0.065} & \multirow{2}{*}{68.52} & 0.23 * & \multicolumn{1}{c|}{0.005} & \multirow{2}{*}{66.86} & -15.59 *** & \multicolumn{1}{c|}{0.012} & \multirow{2}{*}{68.29} & 5.65 *** & 0.028 \\
 &  & \multicolumn{1}{c|}{Organizer} &  & 6.86 *** & \multicolumn{1}{c|}{0.064} &  & -1.02 *** & \multicolumn{1}{c|}{0.012} &  & -6.19 *** & \multicolumn{1}{c|}{0.094} &  & 6.25 *** & 0.049 \\
 & \multirow{2}{*}{Developer} & \multicolumn{1}{c|}{End-User} & \multirow{2}{*}{78.3} & 7.61 *** & \multicolumn{1}{c|}{0.065} & \multirow{2}{*}{68.29} & -0.23 * & \multicolumn{1}{c|}{0.005} & \multirow{2}{*}{82.45} & 15.59 *** & \multicolumn{1}{c|}{0.012} & \multirow{2}{*}{62.64} & -5.65 *** & 0.028 \\
 &  & \multicolumn{1}{c|}{Organizer} &  & 14.47 *** & \multicolumn{1}{c|}{0.136} &  & -1.25 & \multicolumn{1}{c|}{-} &  & 9.4 ** & \multicolumn{1}{c|}{0.009} &  & 0.6 & - \\
 & \multirow{2}{*}{Organizer} & \multicolumn{1}{c|}{End-User} & \multirow{2}{*}{63.83} & -6.86 *** & \multicolumn{1}{c|}{0.064} & \multirow{2}{*}{69.54} & 1.02 *** & \multicolumn{1}{c|}{0.012} & \multirow{2}{*}{73.05} & 6.19 *** & \multicolumn{1}{c|}{0.094} & \multirow{2}{*}{62.04} & -6.25 *** & 0.049 \\
 &  & \multicolumn{1}{c|}{Developer} &  & -14.47 *** & \multicolumn{1}{c|}{0.136} &  & 1.25 & \multicolumn{1}{c|}{-} &  & -9.4 ** & \multicolumn{1}{c|}{0.009} &  & -0.6 & - \\ \hline
\multirow{6}{*}{Clout} & \multirow{2}{*}{End-user} & \multicolumn{1}{c|}{Developer} & \multirow{2}{*}{32.29} & -17.71 *** & \multicolumn{1}{c|}{0.191} & \multirow{2}{*}{32.1} & -17.9 *** & \multicolumn{1}{c|}{0.093} & \multirow{2}{*}{28.91} & -12.54 * & \multicolumn{1}{c|}{0.007} & \multirow{2}{*}{39.7} & -30.54 *** & 0.129 \\
 &  & \multicolumn{1}{c|}{Organizer} &  & -23.23 *** & \multicolumn{1}{c|}{0.300} &  & -17.9 *** & \multicolumn{1}{c|}{0.168} &  & -30.73 *** & \multicolumn{1}{c|}{0.432} &  & -27.11 *** & 0.169 \\
 & \multirow{2}{*}{Developer} & \multicolumn{1}{c|}{End-User} & \multirow{2}{*}{50} & 17.71 *** & \multicolumn{1}{c|}{0.191} & \multirow{2}{*}{50} & 17.9 *** & \multicolumn{1}{c|}{0.093} & \multirow{2}{*}{41.45} & 12.54 * & \multicolumn{1}{c|}{0.007} & \multirow{2}{*}{70.24} & 30.54 *** & 0.129 \\
 &  & \multicolumn{1}{c|}{Organizer} &  & -5.52 *** & \multicolumn{1}{c|}{0.019} &  & 0 *** & \multicolumn{1}{c|}{0.017} &  & -18.19 *** & \multicolumn{1}{c|}{0.015} &  & 3.43 *** & 0.033 \\
 & \multirow{2}{*}{Organizer} & \multicolumn{1}{c|}{End-User} & \multirow{2}{*}{55.52} & 23.23 *** & \multicolumn{1}{c|}{0.300} & \multirow{2}{*}{50} & 17.9 *** & \multicolumn{1}{c|}{0.168} & \multirow{2}{*}{59.64} & 30.73 *** & \multicolumn{1}{c|}{0.432} & \multirow{2}{*}{66.81} & 27.11 *** & 0.169 \\
 &  & \multicolumn{1}{c|}{Developer} &  & 5.52 *** & \multicolumn{1}{c|}{0.019} &  & 0 *** & \multicolumn{1}{c|}{0.017} &  & 18.19 *** & \multicolumn{1}{c|}{0.015} &  & -3.43 *** & 0.033 \\ \hline
\multirow{6}{*}{Tone} & \multirow{2}{*}{End-user} & \multicolumn{1}{c|}{Developer} & \multirow{2}{*}{46.47} & 20.7 *** & \multicolumn{1}{c|}{0.058} & \multirow{2}{*}{44.45} & 18.68 & \multicolumn{1}{c|}{-} & \multirow{2}{*}{51.3} & 5.07 & \multicolumn{1}{c|}{-} & \multirow{2}{*}{54.97} & 5.65 *** & 0.013 \\
 &  & \multicolumn{1}{c|}{Organizer} &  & 20.7 *** & \multicolumn{1}{c|}{0.085} &  & 18.68 *** & \multicolumn{1}{c|}{0.113} &  & 25.53 *** & \multicolumn{1}{c|}{0.169} &  & 29.2 *** & 0.082 \\
 & \multirow{2}{*}{Developer} & \multicolumn{1}{c|}{End-User} & \multirow{2}{*}{25.77} & -20.7 *** & \multicolumn{1}{c|}{0.058} & \multirow{2}{*}{25.77} & -18.68 & \multicolumn{1}{c|}{-} & \multirow{2}{*}{46.23} & -5.07 & \multicolumn{1}{c|}{-} & \multirow{2}{*}{49.32} & -5.65 *** & 0.013 \\
 &  & \multicolumn{1}{c|}{Organizer} &  &  & \multicolumn{1}{c|}{-} &  & 0 *** & \multicolumn{1}{c|}{0.103} &  & 20.46 * & \multicolumn{1}{c|}{0.007} &  & 23.55 *** & 0.072 \\
 & \multirow{2}{*}{Organizer} & \multicolumn{1}{c|}{End-User} & \multirow{2}{*}{25.77} & -20.7 *** & \multicolumn{1}{c|}{0.085} & \multirow{2}{*}{25.77} & -18.68 *** & \multicolumn{1}{c|}{0.113} & \multirow{2}{*}{25.77} & -25.53 *** & \multicolumn{1}{c|}{0.169} & \multirow{2}{*}{25.77} & -29.2 *** & 0.082 \\
 &  & \multicolumn{1}{c|}{Developer} &  &  & \multicolumn{1}{c|}{-} &  & 0 *** & \multicolumn{1}{c|}{0.103} &  & -20.46 * & \multicolumn{1}{c|}{0.007} &  & -23.55 *** & 0.072 \\ \hline
\multirow{6}{*}{Authentic} & \multirow{2}{*}{End-user} & \multicolumn{1}{c|}{Developer} & \multirow{2}{*}{38.39} & 19.12 *** & \multicolumn{1}{c|}{0.106} & \multirow{2}{*}{29.23} & 11.77 *** & \multicolumn{1}{c|}{0.039} & \multirow{2}{*}{47.48} & 36.79 *** & \multicolumn{1}{c|}{0.011} & \multirow{2}{*}{35.37} & 16. *** & 0.064 \\
 &  & \multicolumn{1}{c|}{Organizer} &  & 17.92 *** & \multicolumn{1}{c|}{0.143} &  & 21.39 *** & \multicolumn{1}{c|}{0.145} &  & 27.66 *** & \multicolumn{1}{c|}{0.268} &  & 15.27 *** & 0.096 \\
 & \multirow{2}{*}{Developer} & \multicolumn{1}{c|}{End-User} & \multirow{2}{*}{19.27} & -19.12 *** & \multicolumn{1}{c|}{0.106} & \multirow{2}{*}{17.46} & -11.77 *** & \multicolumn{1}{c|}{0.039} & \multirow{2}{*}{10.69} & -36.79 *** & \multicolumn{1}{c|}{0.011} & \multirow{2}{*}{19.37} & -16. *** & 0.064 \\
 &  & \multicolumn{1}{c|}{Organizer} &  & -1.2 *** & \multicolumn{1}{c|}{0.015} &  & 9.62 *** & \multicolumn{1}{c|}{0.080} &  & -9.13 & \multicolumn{1}{c|}{-} &  & -0.73 & - \\
 & \multirow{2}{*}{Organizer} & \multicolumn{1}{c|}{End-User} & \multirow{2}{*}{20.47} & -17.92 *** & \multicolumn{1}{c|}{0.143} & \multirow{2}{*}{7.84} & -21.39 *** & \multicolumn{1}{c|}{0.145} & \multirow{2}{*}{19.82} & -27.66 *** & \multicolumn{1}{c|}{0.268} & \multirow{2}{*}{20.1} & -15.27 *** & 0.096 \\
 &  & \multicolumn{1}{c|}{Developer} &  & 1.2 *** & \multicolumn{1}{c|}{0.015} &  & -9.62 *** & \multicolumn{1}{c|}{0.080} &  & 9.13 & \multicolumn{1}{c|}{-} &  & 0.73 & - \\ \hline
\multirow{6}{*}{cogproc} & \multirow{2}{*}{End-user} & \multicolumn{1}{c|}{Developer} & \multirow{2}{*}{12.31} & -0.37 *** & \multicolumn{1}{c|}{0.025} & \multirow{2}{*}{12.5} & 0.00 & \multicolumn{1}{c|}{-} & \multirow{2}{*}{12.5} & 2.12 & \multicolumn{1}{c|}{-} & \multirow{2}{*}{13.73} & -2.3 *** & 0.065 \\
 &  & \multicolumn{1}{c|}{Organizer} &  & -1.67 *** & \multicolumn{1}{c|}{0.100} &  & -0.12 & \multicolumn{1}{c|}{-} &  & 0 *** & \multicolumn{1}{c|}{0.009} &  & -0.56 *** & 0.019 \\
 & \multirow{2}{*}{Developer} & \multicolumn{1}{c|}{End-User} & \multirow{2}{*}{12.68} & 0.37 *** & \multicolumn{1}{c|}{0.025} & \multirow{2}{*}{12.5} & 0.00 & \multicolumn{1}{c|}{-} & \multirow{2}{*}{10.38} & -2.12 & \multicolumn{1}{c|}{-} & \multirow{2}{*}{16.03} & 2.3 *** & 0.065 \\
 &  & \multicolumn{1}{c|}{Organizer} &  & -1.3 *** & \multicolumn{1}{c|}{0.057} &  & -0.12 & \multicolumn{1}{c|}{-} &  & -2.12 & \multicolumn{1}{c|}{-} &  & 1.74 *** & 0.103 \\
 & \multirow{2}{*}{Organizer} & \multicolumn{1}{c|}{End-User} & \multirow{2}{*}{13.98} & 1.67 *** & \multicolumn{1}{c|}{0.100} & \multirow{2}{*}{12.61} & 0.12 & \multicolumn{1}{c|}{-} & \multirow{2}{*}{12.5} & 0 *** & \multicolumn{1}{c|}{0.009} & \multirow{2}{*}{14.29} & 0.56 *** & 0.019 \\
 &  & \multicolumn{1}{c|}{Developer} &  & 1.3 *** & \multicolumn{1}{c|}{0.057} &  & 0.12 & \multicolumn{1}{c|}{-} &  & 2.12 & \multicolumn{1}{c|}{-} &  & -1.74 *** & 0.103 \\ \hline
\multirow{6}{*}{affect} & \multirow{2}{*}{End-user} & \multicolumn{1}{c|}{Developer} & \multirow{2}{*}{3.32} & 1.19 *** & \multicolumn{1}{c|}{0.110} & \multirow{2}{*}{3.47} & 0.16 *** & \multicolumn{1}{c|}{0.015} & \multirow{2}{*}{3.57} & 1.49 *** & \multicolumn{1}{c|}{0.009} & \multirow{2}{*}{3.61} & 0.43 *** & 0.027 \\
 &  & \multicolumn{1}{c|}{Organizer} &  & 0.84 *** & \multicolumn{1}{c|}{0.125} &  & 3.47 *** & \multicolumn{1}{c|}{0.176} &  & 1.01 *** & \multicolumn{1}{c|}{0.181} &  & 0.91 *** & 0.096 \\
 & \multirow{2}{*}{Developer} & \multicolumn{1}{c|}{End-User} & \multirow{2}{*}{2.13} & -1.19 *** & \multicolumn{1}{c|}{0.110} & \multirow{2}{*}{3.31} & -0.16 *** & \multicolumn{1}{c|}{0.015} & \multirow{2}{*}{2.08} & -1.49 *** & \multicolumn{1}{c|}{0.009} & \multirow{2}{*}{3.18} & -0.43 *** & 0.027 \\
 &  & \multicolumn{1}{c|}{Organizer} &  & -0.35 *** & \multicolumn{1}{c|}{0.035} &  & 3.31 *** & \multicolumn{1}{c|}{0.132} &  & -0.48 & \multicolumn{1}{c|}{-} &  & 0.48 *** & 0.06 \\
 & \multirow{2}{*}{Organizer} & \multicolumn{1}{c|}{End-User} & \multirow{2}{*}{2.48} & -0.84 *** & \multicolumn{1}{c|}{0.125} & \multirow{2}{*}{0} & -3.47 *** & \multicolumn{1}{c|}{0.176} & \multirow{2}{*}{2.56} & -1.01 *** & \multicolumn{1}{c|}{0.181} & \multirow{2}{*}{2.7} & -0.91 *** & 0.096 \\
 &  & \multicolumn{1}{c|}{Developer} &  & 0.35 *** & \multicolumn{1}{c|}{0.035} &  & -3.31 *** & \multicolumn{1}{c|}{0.132} &  & 0.48 & \multicolumn{1}{c|}{-} &  & -0.48 *** & 0.06 \\ \hline
\multirow{6}{*}{drives} & \multirow{2}{*}{End-user} & \multicolumn{1}{c|}{Developer} & \multirow{2}{*}{5.13} & 0.78 *** & \multicolumn{1}{c|}{0.054} & \multirow{2}{*}{5.07} & 0.31 *** & \multicolumn{1}{c|}{0.014} & \multirow{2}{*}{4.69} & 1.32 *** & \multicolumn{1}{c|}{0.009} & \multirow{2}{*}{5.88} & 0.32 & - \\
 &  & \multicolumn{1}{c|}{Organizer} &  & 0.31 *** & \multicolumn{1}{c|}{0.036} &  & 5.07 *** & \multicolumn{1}{c|}{0.151} &  & 0.29 *** & \multicolumn{1}{c|}{0.043} &  & 0.62 *** & 0.042 \\
 & \multirow{2}{*}{Developer} & \multicolumn{1}{c|}{End-User} & \multirow{2}{*}{4.35} & -0.78 *** & \multicolumn{1}{c|}{0.054} & \multirow{2}{*}{4.76} & -0.31 *** & \multicolumn{1}{c|}{0.014} & \multirow{2}{*}{3.37} & -1.32 *** & \multicolumn{1}{c|}{0.009} & \multirow{2}{*}{5.56} & -0.32 & - \\
 &  & \multicolumn{1}{c|}{Organizer} &  & -0.47 *** & \multicolumn{1}{c|}{0.039} &  & 4.76 *** & \multicolumn{1}{c|}{0.117} &  & -1.03 ** & \multicolumn{1}{c|}{0.008} &  & 0.3 *** & 0.039 \\
 & \multirow{2}{*}{Organizer} & \multicolumn{1}{c|}{End-User} & \multirow{2}{*}{4.82} & -0.31 *** & \multicolumn{1}{c|}{0.036} & \multirow{2}{*}{0} & -5.07 *** & \multicolumn{1}{c|}{0.151} & \multirow{2}{*}{4.4} & -0.29 *** & \multicolumn{1}{c|}{0.043} & \multirow{2}{*}{5.26} & -0.62 *** & 0.042 \\
 &  & \multicolumn{1}{c|}{Developer} &  & 0.47 *** & \multicolumn{1}{c|}{0.039} &  & -4.76 *** & \multicolumn{1}{c|}{0.117} &  & 1.03 ** & \multicolumn{1}{c|}{0.008} &  & -0.3 *** & 0.039 \\ \hline

\end{tabular}%
}

\label{tab:post-hoc-results}
\end{table*}


For the `Analytic' category, the difference was significant for all pairs of user groups on the Zotero and Audacity forums, while only the differences between end-users and developers and end-users and organizers were significant for VLC and RStudio. Particularly, developers were the most analytical on the Zotero and Audacity forums, while on RStudio and VLC, organizers and end-users were the most analytical respectively. 
Moreover, Zotero's median analytical scores for end-users were higher than organizers while they were opposite in Audacity.

\textbf{\textit{Insight-Analytic}. The use of analytical speech among forum user groups appeared to be forum-dependent.}


On the dimension of `Clout', the scores were fairly consistent throughout the forums. All forums saw significant differences among all their participant pairs, with
 end-users' median `Clout' scores the lowest, indicating end-users' conversations lacked confidence. 
Zotero and Audacity's median organizer scores were higher than the developer's scores, VLC had equal median scores for organizers and developers, and RStudio experienced median developer scores higher than median organizer scores.

\textbf{\textit{Insight-Clout.} End-users exhibited the lowest clout scores (i.e., confidence) than the other two groups in all forums.} 


The tonality of the participants' posts across the forums illustrated that the differences between end-users and organizers are the most significant;
end-users exhibited higher median scores than organizers for all forums and, in cases where end-users were significantly different from developers (Zotero, VLC, and RStudio), were also higher than developers. The difference between developers and organizers was significant for three of the forums (Audacity, RStudio, and VLC) but there were conflicting results on which groups have higher median scores. 
Lastly, all median scores for end-users (in addition to the developers of Audacity and RStudio) were near 50 (neutral tones) while the median scores for all organizers and the remaining developer groups were 25.77 (negative tones).

\textbf{\textit{Insight-Tone.} End-users exhibited significantly higher median scores (i.e. more positive with respect to developers and organizers, but still neutral in tone) than organizers for all forums and developers for Zotero and VLC.}


The `Authentic' category results for the four forums illustrate that end-users were significantly more honest and open (with the highest median scores) than the
organizers and developers across all forums.
However, only Zotero and VLC had a significant difference between developers and organizers. 
The developer-organizer pair in Zotero had
a median developer score higher than the organizer. Oppositely, the developer-organizer pair in VLC had
a median organizer score higher than the developer. 

\textbf{\textit{Insight-Authentic.} End-users exhibited more openness a-\\nd honesty than organizers and developers in all forums.}


Moving to the dictionary categories, we inspected the cognitive processes. Similar to `Authentic', the `cogproc' category exhibited quite diverse results: (a) VLC had no significant pairs; (b) Audacity only had one significant pair (EU-ORG); 
(c) both Zotero and RStudio saw all the pairs significantly different.
Further inspection showed significant differences in the three forums (Zotero, RStudio, and Audacity), end-users in Zotero and RStudio had lower median `cogproc' scores than organizers, but EU had the same median scores as organizers in Audacity. 

\textbf{\textit{Insight-cogproc.} The amount of cognitive processes in po-\\sts appeared to be forum-dependent,
yet end-users' posts reflected less cognitive complexity, in general.}


The `affect' category was significantly different for all pairs in three of the forums (Zotero, VLC, RStudio) 
while the Audacity forum was not different between developers and organizers. For all four forums, the end-users
had a median higher than organizers and developers.
Moreover, an inspection into the subcategories of `posemo' and `negemo' showed more use of positive emotion words than the negative emotion words despite experiencing issues or negative occurrences with the OSS that led them to use the forum. 

\textbf{\textit{Insight-affect.} End-users had consistently different and h-\\igher scores than organizers and developers in all forums.}


Last, the `drives' category experienced consistent differences between end-users and organizers where
the median end-user score was higher than the median organizer score.
Moreover, the remaining forums' pairs (end-users and developers, developers and organizers) were all significantly different except for the RStudio's end-user and developer pair. 
Additionally, the developer and organizer median scores were inconsistent; Zotero and Audacity's median organizer scores were higher than the developers' while their scores reversed in RStudio and VLC.

\textbf{\textit{Insight-drives.} End-users appeared to be more need-driv-\\en and motivated than the other groups, and the developers' and organizers' linguistic behavior was forum-dependent.}

\section{Discussion}
\label{sec:disc}

In this study, we characterized user behaviors in OSS user forums. Section \ref{sec:frequency-participation-results} revealed how different categories of forum users made contributions to OSS user forums following distinct patterns. All forums possessed a large end-user base who not only initiated almost all threads but also replied to other users' posts and displayed more positivity and openness in communication. These findings suggest that this large body of enthusiastic, motivated people could become an asset in the OSS development process by providing firsthand usability and accessibility-related feedback. Moreover, our results indicated that the discussion threads on user forums were often short in time span ($median=17.72$ hours) and the number of comments ($median=3$). Interestingly, forum users were generally responsive in replying to new posts (\textit{median delay} $=10.83$ hours) but they often ignored or abandoned older posts, including their own posts -- perhaps because recent posts are usually displayed first in the forums, while revisiting older posts requires a laborious process (i.e., manually search). Another reason could be the lack of proper guidance on using forums (e.g., properly closing a thread or using forum features) that we observed during manual inspection of the forums. Therefore, further work is required to explore ways to aid end-users' on-boarding process to the OSS community.  

The insights from linguistic analyses (Section \ref{sec:liwc-results}) present a foundation for identifying potential strategies moving forward to encourage end-users to participate in the OSS development process. In particular, the insights from the `Clout', `Tone', `Authentic', and `Affect' categories show that end-users in all four forums are significantly less confident than the organizers, but are simultaneously more divulging, open, and positive in their communications. Since the current OSS development process often overlooks the contributions from end-users~\cite{wang_how_2020, hellman_facilitating_2021}, some end-users can be reluctant or apprehensive to use forums unless it is a last resort for obtaining help \cite{hellman_facilitating_2021}. This is likely reflected in the end-user LIWC results with the low `Clout' scores but higher `Tone', `Authentic', and `affect'. 
Towards the goal of bringing end-users who may not be able to contribute codes into the OSS development pipeline, the forum end-users' inclination towards more neutral or positive, personable, analytical communication is promising. Especially when combined with the fact that the end-users used the most words in the 'drives' category (indicating they were working towards a specific goal to achieve), there is support to our claim that some end-users from the vast forum end-user base would be interested in participating further in the OSS development process, though they lack the confidence to do so now. 

The LIWC analysis results indicated that organizers and developers could make more efforts through their language. Based on these findings, we provide the following recommendation to better support helping and sustaining end-users: We suggest for organizers and developers to adopt a more open, personal, and positive tone to match the current practice of end-users when communicating in the forums. If organizers and developers can become more welcoming and supportive of end-users in the forums, then ideally it can encourage more end-users to participate in various aspects of the OSS development processes, even if they lack the skills to code. Looking to the future, including end-users in the OSS development process will be essential for improving OSS, in particular as end-user bases grow and diversify~\cite{wang_how_2020}. 

Finally, the `Analytic' category was one of a few whose results indicated features unique to the specific forums. While we expected that developers or organizers would be most analytical in their speech based on the assumption that these stakeholders would use logical arguments and explanations to help end-users, the results did not support this hypothesis; we found that, in some forums, developers were more analytical than end-users but was the opposite in others. 
The cause of these phenomena might be related to the level of end-user participation in investigating and answering questions posted by other users. For example, RStudio and VLC had more \textit{replies by others} performed by end-users than developers and organizers, corresponding with the higher average analytical scores for end-users. 
While further analysis is needed to confirm this as the cause for the `Analytic' scores, these results illustrate that for this category, a forum's participant groups can exhibit different behaviors that are dependent on the context of the forum itself.

\section{Limitations \& Threats to Validity}
This study suffers from the following limitations and threats to validity that we plan to address in the future. 

First, we were only able to focus on four OSS user forums in our analysis. Our sample was selected from the Open Source Software Directory~\footnote{opensourcesoftwaredirectory.com}, across different categories and sorted alphabetically, where we targeted the most downloaded OSS applications that the authors were already familiar with. Moreover, while all the forums that were selected are managed in English, occasionally users wrote their posts in other languages (around 4\% per forum based on our investigations). Initial investigations with automatic language detection methods experienced many false negatives; therefore we decided to process all posts as the non-English posts only trivially impacted the results.

Second, when identifying the roles of forum users, we used their user ID as a token to cross-reference developers from the GitHub repository of the OSS projects. While our manual inspection indicated that developers tended to use the same user ID on both the forums and GitHub, we cannot guarantee that this cross-reference is complete. We found that the percentages of GitHub contributors matched with the forum users of the Zotero, Audacity, VLC, and RStudio projects are, respectively: 42.1\%; 3.1\%; 6.5\%; and 22.6\%.
Furthermore, we considered the end-users as one single group despite our analysis revealing a wide range in the demonstrated activities (i.e., some users have participated since project inception while most participated only sparsely).

Third, our analysis adopted mostly a quantitative methodology which allowed us to manage a large amount of data;
future studies focusing on qualitative analysis of a sample of forum discussions may help explain more of our insights. Particularly, in regards to thread topics and power dynamics between user groups.

Finally, our analysis is based only on the observed behavior of the user forums. We acknowledge that a large amount of end-users of OSS projects are not able to or choose not to contribute and engage in the user forums. Their perspectives and experiences with the OSS application, although important, are not visible. Future work focusing on user studies, such as surveys and interviews with OSS end-users, would be useful to understand these and other invisible perspectives.

\section{Conclusion}
We performed an empirical study on over 1.3 million posts from four popular OSS user forums: Zotero, Audacity, VLC, and RStudio. Our study employed two methods (1) analyzing forum users' roles and types of activities, particularly the frequency and breadth of user interactions, and (2) conducting psycho-linguistic analysis using LIWC to analyze the contribution patterns and behaviors amongst the forum user categories (end-users, developers, and organizers). We found that these three user categories can make an initial post, reply as the original poster, or reply as another poster. End-users make the most initial posts but also actively engage in other peoples' thread conversations. Moreover, the LIWC results show significantly different linguistic behavior between user pairs, particularly in terms of `Clout', `Tone', `Authentic', and `affect'.  Overall, end-users demonstrate more positive and disclosing natures compared to developers and organizers in forums but are significantly less confident. Our work presents novel characterizations of user behaviors in OSS user forums to be leveraged to increase end-user engagement in the larger OSS community.

\begin{acks}
This work is partially supported by Alfred P. Sloan Foundation (Grant No.:G-2021-16745). We also thank the anonymous reviewers for their valuable feedback.
\end{acks}

\bibliographystyle{ACM-Reference-Format}
\bibliography{ref}

\end{document}